\documentclass[12pt]{iopart}
\usepackage{graphicx}

\newcommand{\mb}[1]{\mbox{\boldmath $#1$}}
\def \ts {\textstyle}
\def\be{\begin{equation}}
\def\ee{\end{equation}}
\def\beq{\begin{eqnarray}}
\def\eeq{\end{eqnarray}}

\newcommand{\grqc}[1]{arXiv:gr-qc/#1}
\newcommand{\astroph}[1]{arXiv:astro-ph/#1}

\def\AP{Adv. Phys. }
\def\ApJ{Astrophys. J. }
\def\AWL{Akad. Wiss. Lit. Mainz. Abhandl. Math.-Nat. Kl. }
\def\CMP{Commun. Math. Phys. }
\def\CQG{Class. Quant. Grav. }
\def\ETF{Eksp. Teor. Fiz. }
\def\GRG{Gen. Relativ. Gravit. }
\def\JETP{Sov. Phys. JETP }
\def\PRD{Phys. Rev. D  }
\def\PRL{Phys. Rev. Lett. }
\def\PTP{Prog. Theor. Phys. }
\begin{document}
\jl{6}

\title[Covariant Fluid Dynamics: a Long Wave-Length Approximation]
{Covariant Fluid Dynamics: a Long Wave-Length Approximation}

\author{Marco Bruni and Carlos F. Sopuerta\ddag}
\address{Institute of Cosmology and Gravitation, Mercantile House,
University of Portsmouth, Portsmouth PO1 2EG, Britain}

\date{\today}

%
%

\begin{abstract}
In this paper we consider the Long-Wavelength Approximation Scheme (LWAS)
in the framework of the covariant fluid approach to general
relativistic dynamics, specializing to the particular case of irrotational
dust matter.  We discuss the dynamics of these models during the approach to
any spacelike singularity where a BKL-type evolution is expected, studying the
validity of this approximation scheme and the role of  the magnetic part of the
Weyl tensor, $H_{ab}$.  Our analytic results confirm a previous numerical
analysis: it is $H_{ab}$ that destroys the pure Kasner-like approach
to the singularity and eventually produces the bounce to another Kasner
phase.  Expanding regions evolve as separate universes where inhomogeneities
and anisotropies die away.
\end{abstract}




\section{Introduction}\label{intr}
Approximations methods are and have been very important in general relativity
and its applications to cosmology and relativistic astrophysics.  In
the case of cosmology linear perturbation theory~\cite{LIKA,LIPT} is at the heart of
the present developments, in particular, in the analysis of the observed
anisotropies of the Cosmic Microwave Background (see, e.g.,~\cite{BOOK}).
Another approximation technique widely used in cosmology is the so-called
Long Wave-Length Approximation Scheme (LWAS).  Broadly speaking, it
consists in neglecting spatial gradients of the variables
describing the cosmological models, assuming they are small in comparison
with their time derivatives.  Since the time-scale of variation in cosmology
is given by the local Hubble expansion rate, the approximation consists in
neglecting inhomogeneities varying over a scale smaller than the
local Hubble horizon.  One can then try to get information about scales
smaller that the Hubble radius by computing their effects
through a series expansion in spatial gradients, which constitutes the
LWAS (sometimes also called the gradient expansion method or
long wavelength iteration).
This scheme is then suitable to study large-scale structure formation
and issues related with it.
One important advantage of using the LWAS is the fact that neglecting
spatial gradients we are still taking into account the non-linear character
of the dynamical equations. Actually, the dynamical evolution that
comes out will be fully non-linear and essentially the same as the one for
homogeneous cosmological models.

In the literature there have been several approaches to the LWAS:
using this approximation Belinski, Khalatnikov, and Lifshitz (BKL)
studied~\cite{BKLP} (see also~\cite{BKL2}) the general behaviour of
cosmological models near the initial singularity.  In~\cite{LWA1,TODE} it was
used to study several important issues in structure formation, from the
gravitational
instability mechanism to the evolution of the primordial inhomogeneities.
Higher-order expansions in the LWAS and their applications have been developed
in~\cite{LWA2} by using a Hamiltonian approach.   A closely
related iterative method has been worked out in~\cite{LWA3,LADR}.
Another approximation scheme close to the LWAS was presented
in~\cite{LWA4}.

In this paper we study a different formulation of the LWAS
based on the covariant fluid dynamics approach to cosmology (see, e.g.,
\cite{EHLE,COFA,ELVE}).   In this sense our work is complementary to the
perturbative gauge invariant formalism in~\cite{BDE} (see also~\cite{ELVE} and
references therein).
We focus on the study of irrotational dust models
(IDM) with a cosmological constant $\Lambda>0$, and more specifically, on
the dynamical aspects near the
singularities.   One of the main conclusions of this study is
that the LWAS breaks down due to a gravitomagnetic effect.
This points at what it should be expected in the light of the
standard BKL picture:
that in the generic
case the approach to a singularity has a first phase in which the solution
goes into a Kasner phase, after which there is a ``jump'' into a transient
period in which the solution moves away from a Kasner solution, and then
goes into a new Kasner phase.
With our language, and within the LWAS, this jumping is caused by a
gravitomagnetic effect that is necessarily associated with non degenerate
cases (here degenerate will mean two equal shear eigenvalues).
From the point of view of the evolution equations, the LWAS reduces the
quasi-linear Partial Differential Equations (PDEs) system of Einstein's
equations, or equivalently the covariant fluid equations, to Ordinary
Differential Equations (ODEs).  The evolution in this phase proceeds
in a ``silent'' fashion~\cite{BMP,WAEL} approaching a Kasner singularity,
till the effects of the gravito-magnetic Weyl tensor $H_{ab}$ are no
longer negligible and a bounce to a new Kasner phase is produced.

The plan of this paper is the following. In section~\ref{sec2}
we review the fluid dynamical approach to IDMs and choose the
most suitable form of the equation for the application of the
LWAS.  In section~\ref{sec3} we show how to implement the LWAS
within the chosen approach and we present the first-order
solution.  In section~\ref{sec4} we discuss the validity of the
approximation from the point of view of the covariant fluid
dynamics approach.  We end with some remarks and conclusions
in section~\ref{sec5}.

Throughout this paper we will use units in which $8\pi G=c=1$.
Tensorial components will be expressed both in coordinate charts and in
vector basis.  Then, our conventions for indices are:
spacetime coordinate indices are denoted by the lower-case Latin letters
$a,\ldots,l=0,1,2,3$, and spacetime indices associated with an
arbitrary basis $\{\mb{e}_0,\mb{e}_1,\mb{e}_2,\mb{e}_3\}$ by the
rest of lower-case Latin letters $m,\ldots, z=0,1,2,3$.
Indices with respect to an orthonormal triad of spacelike vectors
$\{\mb{e}_1,\mb{e}_2,\mb{e}_3\}$ are denoted by lower-case Greek letters
$\alpha,\ldots, \lambda=1,2,3$ and spatial coordinate indices by
the rest of lower-case Greek letters $\mu,\ldots,\omega=1,2,3$.


\section{Irrotational Dust Models}\label{sec2}

IDMs are important both for the description of the late universe as well
as for the study of the gravitational instability mechanism.
Their energy-momentum distribution is completely described by the
fluid velocity $\mb{u}$, a unit timelike vector field ($u^au_a=-1$),
and its associated energy density $\rho$.  The energy-momentum tensor is given by
\begin{equation}
T_{ab} = \rho  u_au_b \,, \label{emte}
\end{equation}
Since there is no pressure, the energy-momentum conservation equations
tell us that the fluid worldlines are geodesics, i.e., $u^b\nabla_bu^a=0$.
Assuming the fluid flow to be irrotational ($u_{[a}\nabla_bu_{c]}=0$) implies
that the fluid velocity is the normal to a foliation of spacelike
hypersurfaces.  Then, there exists a time function $\tau(x^a)$ such that
$\mb{u}$ is given by
\[ \mb{\vec{u}} = \frac{\mb{\partial}}{\mb{\partial\tau}}\,,
\hspace{10mm} \mb{u} = -\mb{d\tau} \,. \]
That is, $\tau$ is at the same time the proper time
of the fluid elements and a label of the hypersurfaces orthogonal
to $\mb{u}\,,$ which we will denote by $\Sigma(\tau_1)\equiv\{\tau(x^a)=\tau_1:
\mbox{constant}\}\,.$  A system of coordinates for these hypersurfaces
can be formed from any three independent first integrals of $\mb{u}$,
$\{y^\mu(x^a)\}$ with $u^a\partial_ay^\mu=0$.  Then, $\{\tau,y^\mu\}$ constitutes
a set of comoving synchronous geodesic normal coordinates, in which the line element 
takes the following form
\begin{equation}
ds^2 = -d\tau^2 + h_{\mu\nu}(\tau,y^\sigma)dy^\mu dy^\nu \,, \label{liel}
\end{equation}
where $h_{\mu\nu}$ are the nonzero components of the induced positive-definite
metric (first fundamental form) on the hypersurfaces $\Sigma(\tau)$.  This
coordinate system is fixed up to the transformations:
$\tau\,\rightarrow \tau'= \tau+constant$, and
$y^\mu\,\rightarrow\,y^{\mu'}= f^{\mu'}(y^\nu)$, being $f^{\mu'}$
arbitrary functions.

A very convenient description of these models is provided by the
covariant fluid approach introduced by Ehlers~\cite{EHLE}
(see~\cite{COFA,ELVE} for details).  For the purposes of this paper
it will be useful to write the equations using an
orthonormal tetrad adapted to the fluid velocity:
$\{\mb{e}_0=\mb{u},\mb{e}_1,\mb{e}_2,\mb{e}_3\}$ with
$\mb{e}_m\cdot\mb{e}_n=\eta_{mn}\equiv\mbox{diag} (-1,1,1,1)\,.$
For convenience we fix partially the freedom in the choice of the
triad $\{\mb{e}_\alpha\}$ by requiring it to propagate parallely to
the fluid flow
\[ \dot{e}_\alpha{}^a = u^b\nabla_b\,e_\alpha{}^a =0\,. \]
This choice also makes the local angular velocity to vanish
\[ \Omega^\alpha\equiv\textstyle{1\over2}\varepsilon^{\alpha\beta
\delta}\mb{e}_\beta\cdot\dot{\mb{e}}_\delta = 0~~~~~
(\varepsilon_{\alpha\beta\delta}=\eta_{\alpha\beta\delta 0})\,,\]
where $\eta_{abcd}$ is the spacetime volume element.  In this
framework, the variables that we need to describe IDMs are the 
following.
(i) {\it Metric variables}. The components of the triad vectors
in adapted coordinates $\{\tau,y^\mu\}$, $e_\alpha{}^\mu$.
(ii) {\it Connection variables}.
The spatial commutators, $\gamma^\alpha{}_{\beta\lambda}$, defined by
the commutation relations between the triad vectors:
\begin{equation}
\left[\mb{e}_\beta,\mb{e}_\lambda\right]=
\gamma^\alpha{}_{\beta\lambda}\mb{e}_\alpha\,,\hspace{5mm}
\gamma^\alpha{}_{[\beta\lambda]}=\gamma^\alpha{}_{\beta\lambda}\,.
\label{spco}
\end{equation}
We can instead use, the following variables introduced by Sch\"ucking, Kundt
and Behr (see, e.g.,~\cite{ELMC}):
\begin{equation}
\gamma^\alpha{}_{\beta\lambda} = 2 a_{[\beta}\delta^\alpha{}_{\lambda]}
+\varepsilon_{\beta\lambda\delta}n^{\alpha\delta}~~
\Longleftrightarrow ~~
 a_\alpha=\textstyle{1\over2}\gamma^\beta{}_{\alpha\beta} \,,
\hspace{4mm} n^{\alpha\beta}=\textstyle{1\over2}
\varepsilon^{\lambda\delta(\alpha}\gamma^{\beta)}{}_{\lambda\delta}\,,
\label{coco}
\end{equation}
which contain the same information as $\gamma^\alpha{}_{\beta\lambda}$.
(iii) {\it Kinematical variables}. The expansion $\Theta$
($\equiv\nabla_au^a$) and the shear tensor of the fluid worldlines.
The shear is a symmetric and trace-free {\em spatial}\footnote{{\em Spatial}
means orthogonal to the fluid velocity.} tensor
\[ \sigma_{ab}\equiv h_{(a}{}^ch_{b)}{}^d\nabla_du_c-(\Theta/3)h_{ab}
= \nabla_au_b-(\Theta/3)h_{ab} \,. \]
(iv) {\em Matter variables}. The only non-zero component of the
energy-momentum tensor is the energy density $\rho= T_{00}$.  The Ricci
tensor is, through Einstein's equations, given by
\begin{equation}
R_{ab}= \rho\,u_au_b + \textstyle{1\over2}(\rho+4\Lambda)g_{ab} \,,
\label{eins}
\end{equation}
where $\Lambda$ denotes the cosmological constant.
(v) {\em Weyl
tensor variables}.  The Weyl tensor $C_{abcd}$ describes
the spacetime curvature not determined locally by matter fields.
Its ten independent components can be divided into two spatial,
symmetric and trace-free tensors
\[ E_{ab}=C_{acbd}u^cu^d\,,\hspace{4mm}
H_{ab}=*C_{acbd}u^cu^d\hspace{2mm} (*C_{abcd}\equiv
\textstyle{1\over2}\eta_{ab}{}^{ef}C_{cdef}) \,, \]
which are called the gravito-electric and gravito-magnetic fields 
respectively. Whereas the
gravito-electric field produces tidal forces, having a Newtonian
analogue (the trace-free part of the Hessian of the Newtonian
potential), the gravito-magnetic field has no Newtonian
analogue.

The equations governing the behaviour of these quantities come
from the Ricci identities applied to $\mb{u}$, the second Bianchi
identities, Einstein's equations~(\ref{eins}) and the Gauss-Codazzi equations
(see~\cite{COFA,ELVE} for details). To obtain them we have to split
the covariant derivative $\nabla_a$ into a time derivative along
the fluid worldlines,
$\dot{A}^{a\cdots}{}_{b\cdots}\equiv u^c\nabla_c A^{a\cdots}{}_{b\cdots}$,
and the induced covariant derivative on the hypersurfaces
$\Sigma(\tau)$, $D_c A^{a\cdots}{}_{b\cdots}\equiv h^a{}_e\cdots h^f{}_b
h_c{}^d\nabla_d A^{e\cdots}{}_{f\cdots}$, where $A^{a\cdots}{}_{b\cdots}$
is any arbitrary tensor field.  To simplify the equation that will appear
in this paper it is convenient to
introduce the spatial divergence and curl of an arbitrary 2-index
symmetric tensor\footnote{These definitions are
analogous to those for vector fields:
$\mbox{div}(A)\equiv D_aA^a$ and
$\mbox{curl}A_a\equiv\varepsilon_{abc}D^bA^c\,.$} $A_{ab}$
(see, e.g.,~\cite{MAAR})
\[ \mbox{div}\,(A)_a\equiv D^bA_{ab}\,,\hspace{10mm}
\mbox{curl}\, A_{ab}\equiv\varepsilon_{cd(a}D^cA_{b)}{}^d \,, \]
where $\varepsilon_{abc}$ ($\varepsilon^{abc}
\varepsilon_{def}=3! h^{[a}{}_dh^b{}_eh^{c]}{}_f$) corresponds to the volume
element of the hypersurfaces $\Sigma(\tau)$.  The projection of
these definitions onto a triad $\{\mb{e}_\alpha\}$ gives\footnote{Angled
brackets on indices denote the spatially projected, symmetric and
tracefree part: $A_{\langle\alpha\beta\rangle}=A_{(\alpha\beta)}-
(A^\lambda{}_\lambda/3)\delta_{\alpha\beta}\,.$}
\[ \mbox{curl}(A)_{\alpha\beta}= \varepsilon^{\lambda\delta}
{}_{<\alpha}(\mb{\partial}_{|\lambda|}-a_{|\lambda|})A_{\beta>\delta}
+\textstyle{1\over2}n^\delta{}_\delta A_{\alpha\beta}-
3n_{<\alpha}{}^\delta A_{\beta>\delta} \,, \]
\[ \mbox{div}(A)_\alpha = (\mb{\partial}_\delta-3a_\delta)A^\delta
{}_\alpha-\varepsilon_\alpha{}^{\beta\delta}n_\beta{}^\lambda
A_{\lambda\delta} \,, \]
where $\mb{\partial}_\alpha\equiv e_\alpha{}^\mu\partial_{y^\mu}\,.$
Finally, for two arbitrary spatial symmetric tensors, $A_{ab}$ and $B_{ab}$,
we define the commutator as
\[ [A,B]_{ab}\equiv 2A_{[a}{}^cB_{b]c} \,, \hspace{10mm} [A,B]_a\equiv
\textstyle{1\over2}\varepsilon_{abc}[A,B]^{bc}=\varepsilon_{abc}A^b{}_d
B^{cd} \,. \]

The resulting set of covariant equations is usually divided into two
groups: {\em Evolutions equations}, giving the rate of change of our
quantities along the fluid world-lines, and {\em constraint equations},
relations containing spatial derivatives only.  The explicit form of these
equations has been given in many places (see, e.g.,~\cite{EHLE,COFA,ELVE}).
In the case of IDMs they can be found in~\cite{SOPL,SOMA}.
In order to set up a suitable framework for the application of the LWAS, here 
we will look at these equations from a different point of view than the usual one.
We consider as {\em main} variables the set $\{e_{\alpha}{}^\mu,
\Theta,\sigma_{\alpha\beta}\}$, which will be the dynamical variables.
Their evolution equations can be written as follows:
\begin{equation}
\dot{e}_\alpha{}^\mu = -\left(\textstyle{1\over3}\Theta
\delta_\alpha{}^\beta+\sigma_\alpha{}^\beta\right)
e_\beta{}^\mu\,, \label{base}
\end{equation}
\begin{equation}
\dot{\Theta}= -\textstyle{1\over2}\Theta^2-\textstyle{3\over4}
\sigma^{\alpha\beta}\sigma_{\alpha\beta}+\textstyle{3\over2}\Lambda-
\textstyle{1\over4}{}^3R \,, \label{expa}
\end{equation}
\begin{equation}
\dot{\sigma}_{\alpha\beta}=-\Theta\sigma_{\alpha\beta}
-{}^3S_{\alpha\beta} \,.
\label{shea}
\end{equation}
In these equations, ${}^3R$ and ${}^3S_{\alpha\beta}$ denote
the scalar curvature and the trace-free part of the Ricci tensor of the
hypersurfaces $\Sigma(\tau)$ respectively.  They must be understood as
given in terms of $\gamma^\delta_{\alpha\beta}$ and their derivatives,
through the expression
\[ {}^3R_{\alpha\beta}=
\mb{\partial}_\lambda(\Gamma^\lambda_{\alpha\beta})-
\mb{\partial}_\beta(\Gamma^\lambda_{\alpha\lambda})+
\Gamma^\epsilon_{\alpha\beta}\Gamma^\lambda_{\epsilon\lambda}-
\Gamma^\epsilon_{\alpha\lambda}\Gamma^\lambda_{\epsilon\beta}+
\Gamma^\lambda_{\alpha\epsilon}\gamma^\epsilon_{\beta\lambda}\,, \]
where $\Gamma^\alpha_{\beta\delta}\equiv\mb{e}^\alpha\cdot
(\nabla^{}_{\mb{e}_\delta}\mb{e}_\beta)$ are the Ricci rotation
coefficients associated with the triad $\{\mb{e}_\alpha\}$,
related to $\gamma^\alpha_{\beta\delta}$ by
$\delta_{\alpha\epsilon}\Gamma^\epsilon_{\beta\lambda} =
\delta_{\alpha\epsilon}\gamma^\epsilon_{\lambda\beta}+
\delta_{\beta\epsilon}\gamma^\epsilon_{\alpha\lambda}+
\delta_{\lambda\epsilon}\gamma^\epsilon_{\alpha\beta}\,.$
The remaining variables, $\{\gamma^\delta_{\alpha\beta},\rho,E_{\alpha\beta},
H_{\alpha\beta}\}$, which will be called {\em auxiliary} variables, can
be found in terms of the main ones, either from their definitions
or from the constraints.  The expressions are:

\begin{equation}
\gamma^\delta_{\alpha\beta}\mb{\partial}^{}_\delta y^\mu =
[\mb{\partial}_\alpha,\mb{\partial}_\beta]y^\mu\,, \label{gamm}
\end{equation}
\begin{equation}
\rho = \textstyle{1\over3}\Theta^2-\textstyle{1\over2}
\sigma^{\alpha\beta}\sigma_{\alpha\beta} - \Lambda
+\textstyle{1\over2}{}^3 R \,, \label{li7}
\end{equation}
\begin{equation}
E_{\alpha\beta} = \textstyle{1\over3}\Theta\sigma_{\alpha\beta}-
\sigma_{<\alpha}{}^\delta\sigma_{\beta>\delta}+{}^3S_{\alpha\beta}
\,. \label{li8}
\end{equation}
\begin{equation}
H_{\alpha\beta} = \mbox{curl}(\sigma)_{\alpha\beta}\,. \label{magw}
\end{equation}
These quantities are subject to the following constraint:
\begin{equation}
\mbox{div}(\sigma)_\alpha-\textstyle{2\over3}
\mb{\partial}_\alpha\Theta = 0 \,. \label{monc}
\end{equation}
The remaining equations involved in this formalism can be derived from
those here presented.
It should be noticed that Einstein's equations enter the scheme above
only through equation~(\ref{li7}), through the particular way $\rho$
is related to the geometrical and kinematical variables.


\section{Formulation of the Long Wave-Length Approximation Scheme:
the first-order solution}\label{sec3}

As we have already mentioned the LWAS consists, roughly speaking, in neglecting
spatial gradients with respect to time derivatives.
One can find in the literature several ways of implementing the
LWAS~\cite{LWA1,LWA2,LWA3}.
In the case of IDMs, the spatial gradients correspond to derivatives tangent
to an intrinsically characterized foliation, namely, the one formed by hypersurfaces
orthogonal to the fluid velocity.  In this sense, this approximation scheme is
physically well-motivated for the case of IDMs.  Based on this fact, here we will
adopt a new point of view
based on the covariant fluid approach introduced in the previous section,
which will allow us to clarify several geometrical and physical aspects
of the LWAS.
Although the formalism we develop can be applied in different contexts,
here we focus in studying the approach to spacelike singularities,
complementing in this way the work done by Langlois and
Deruelle~\cite{LADR}.

Let us consider the dynamical equations governing the
behaviour of IDMs shown in the previous section.  
In order to apply the LWAS ideas it is very important to realise that in 
the evolution equations for
the main quantities, Eqs.~(\ref{base})-(\ref{shea}), only second-order
spatial gradients appear, and they are encoded in the spatial curvature
tensor, or more specifically, in ${}^3R$ and ${}^3S_{\alpha\beta}\,.$
Therefore, the zero- and first-order solutions can be found in one
iteration (we will separate them later), which means that we can look directly 
for the first order
solution just by neglecting in these equations terms of order greater than
one in the spatial gradients.  In our case, this means to neglect
the spatial curvature:
\begin{equation}
{}^3R \approx 0\,, \hspace{8mm} {}^3S_{ab} \approx 0 \,, \label{spcu}
\end{equation}
which implies, through Eq.~(\ref{magw}), that
\begin{equation}
D_aH_{bc}\approx 0 \hspace{3mm} \Longrightarrow \hspace{3mm}
\mbox{div}(H)_a \approx 0 \hspace{3mm} \mbox{and} \hspace{3mm}
\mbox{curl}H_{ab} \approx 0 \,. \label{spgm}
\end{equation}
The first important consequence of~(\ref{spcu}) is that from the
evolution equations~(\ref{base})-(\ref{shea}) we obtain a closed
system of equations for $\Theta$ and $\sigma_{\alpha\beta}\,.$
\begin{equation}
\dot{\Theta}= -\textstyle{1\over2}\Theta^2-\textstyle{3\over4}
\sigma^{\alpha\beta}\sigma_{\alpha\beta}+
\textstyle{3\over2}\Lambda \,, \label{evl1}
\end{equation}
\begin{equation}
\dot{\sigma}_{\alpha\beta}= -\Theta\sigma_{\alpha\beta}\,. \label{evl2}
\end{equation}
Second, the expressions for $\rho$ and $E_{\alpha\beta}$ become
purely local
\begin{equation}
\rho = \textstyle{1\over3}\Theta^2-\textstyle{1\over2}\sigma^{\alpha\beta}
\sigma_{\alpha\beta}-\Lambda \,, \label{exro}
\end{equation}
\begin{equation}
E_{\alpha\beta}=\textstyle{1\over3}\Theta\sigma_{\alpha\beta}-
\sigma_{<\alpha}{}^\lambda\sigma_{\beta>\lambda}\,. \label{gete}
\end{equation}

In order to integrate equations~(\ref{evl1}),(\ref{evl2}) we will
further specialize the triad $\{\mb{e}_\alpha\}$ by taking
into account the special structure of the first-order equations
and the freedom: $\mb{e}_\alpha\rightarrow\mb{e}_{\alpha'}
= {\cal A}_{\alpha'}{}^\alpha\mb{e}_\alpha$ with
$\dot{\cal A}_{\alpha'}{}^\alpha = 0\,.$  We use this freedom to have
an eigenbasis of the shear.  That is,
\begin{equation}
\sigma_{\alpha\beta} = 0 \hspace{4mm}
\mbox{for} \hspace{4mm} \alpha\neq \beta \,, \label{sdia}
\end{equation}
This is possible because at the first order we have
${}^3S_{\alpha\beta}=0\,.$  In~\cite{BARO} (see also~\cite{SILE}),
it was shown that a sufficient condition for this is 
$H_{ab}=0$, and in~\cite{DIVH} this result was extended
to the case in which $H_{ab}$ is transverse,
i.e. $\mbox{div}(H)_a=0$.  We have already seen that this last condition
holds at first order [see Eq.~(\ref{spgm})].
As a consequence of (\ref{sdia}) and (\ref{gete}), $E_{\alpha\beta}$
is also diagonal. Hence $[\sigma,E]_{ab}=0\,.$

We can now solve the equations for the first-order solution of
the LWAS.  The procedure that we will follow is very close to the one
used to find the complete class of IDMs with flat spatial geometry~\cite{SOPL}.
We begin by solving equations (\ref{evl1}),(\ref{evl2}) for the expansion $\Theta$
and the only two independent components of the shear, whose information can be
encoded in the following two quantities~(see, e.g., \cite{WAEL} and references
therein)
\begin{equation}
\sigma_+\equiv -\textstyle{3\over2}\sigma_{11}\,, \hspace{4mm}
\sigma_-\equiv \textstyle{\sqrt{3}\over2}(\sigma_{22}-\sigma_{33})\,.
\label{depm}
\end{equation}
The equations for $(\Theta,\sigma_+,\sigma_-)$ then are
\begin{eqnarray}
\dot{\Theta} = -\textstyle{1\over2}(\Theta^2+\sigma^2_++
\sigma^2_--3\Lambda) \,, \label{theta1} \\
\dot{\sigma}_\pm = -\Theta\sigma_\pm \,, \label{sigma1}
\end{eqnarray}
and the solution can then be expressed as
\begin{equation}
\Theta = \frac{\dot{V}}{V}\,, \hspace{8mm}
\sigma_+ = \frac{\Sigma_+}{V}\,, \hspace{8mm}
\sigma_- = \frac{\Sigma_-}{V}\,, \label{solu}
\end{equation}
where $V=V(\tau,y^\mu)$ denotes the proper volume
associated with the fluid flow of $\mb{u}$ and $\Sigma_\pm$ are scalars
independent of $\tau$.
From~(\ref{theta1},\ref{sigma1}) the equation for $V$ is
\begin{equation}
2V\ddot{V}-\dot{V}^2-3\Lambda V^2 + 3\Sigma^2  = 0\,,~~~~
\Sigma^2\equiv \ts{1\over3}\Sigma^2_+ + \Sigma^2_- \,,
\label{vequ}
\end{equation}
and $\Sigma^2$ is a first integral of the system~(\ref{theta1},\ref{sigma1}).
In order to write the solution of this equation one has to consider
separately the following cases: (i) $\Lambda>0$.  Here, we have to
consider also different cases according to a criterion that involves
$\Lambda$, $\Sigma$, and an integration function that we will call
$M$.  Then, when $M^2-4\Lambda\Sigma^2>0$,
the solution of (\ref{vequ}) can be written as
\begin{eqnarray}
\fl V(t,y^\mu) =
\frac{1}{\Lambda}\sqrt{M^2-4\Lambda\Sigma^2}\;\sinh\left[
\frac{\sqrt{3\Lambda}}{2}\left(\tau-{\cal T}\right) \right]
\sinh\left[
\frac{\sqrt{3\Lambda}}{2}\left(\tau-{\cal T}+\delta{\cal T}\right) \right]
\,, \label{case1}
\end{eqnarray}
where ${\cal T}$ is an arbitrary function of the comoving coordinates
$y^\mu$.  When $M^2-4\Lambda\Sigma^2=0$ we have
\begin{eqnarray}
\fl V(t,y^\mu) =
\frac{M}{2\Lambda}\left(\mbox{e}^{\sqrt{2\Lambda}(\tau-{\cal T})}-1 \right)
\,, \label{case2}
\end{eqnarray}
and finally, when $M^2-4\Lambda\Sigma^2<0$, the solution is:
\begin{eqnarray}
\fl V(t,y^\mu) =
\frac{1}{\Lambda}\sqrt{4\Lambda\Sigma^2-M^2}\;\sinh\left[
\frac{\sqrt{3\Lambda}}{2}\left(\tau-{\cal T}\right)\right]
\cosh\left[
\frac{\sqrt{3\Lambda}}{2}\left(\tau-{\cal T}+\delta{\cal T}\right) \right]
\,. \label{case3}
\end{eqnarray}
Where the function $\delta{\cal T}$, in cases (\ref{case1},\ref{case3}),
is given by
\begin{eqnarray}
\delta{\cal T} = \frac{1}{\sqrt{3\Lambda}}\ln\left|\frac{M+2\sqrt{\Lambda}\Sigma}
{M-2\sqrt{\Lambda}\Sigma} \right| \,.
\end{eqnarray}
(ii) $\Lambda<0$.  In this case the solution for the comoving volume is
\begin{eqnarray}
\fl V(t,y^\mu) =
-\frac{1}{\Lambda}\sqrt{M^2-4\Lambda\Sigma^2}\;\sin\left[
\frac{\sqrt{-3\Lambda}}{2}\left(\tau-{\cal T}\right)\right]
\cos\left[\frac{\sqrt{-3\Lambda}}{2}\left(\tau-{\cal T}+\delta{\cal T}\right)
\right]\,, \nonumber
\end{eqnarray}
where
\begin{eqnarray}
\delta{\cal T} = -\frac{2}{\sqrt{-3\Lambda}}\arcsin\left[
\frac{M}{\sqrt{M^2-4\Lambda\Sigma^2}} \right]\,.
\end{eqnarray}
(iii) $\Lambda=0$.  The solution for this case is:
\begin{eqnarray}
 V(t,y^\mu) = \textstyle{3\over4}M(\tau-{\cal T})(\tau-{\cal T}+\delta{\cal T})\,,
\label{prvo}
\end{eqnarray}
where
\begin{eqnarray}
\hspace{10mm} \delta{\cal T}\equiv\frac{4\Sigma}
{\sqrt{3}M}\,.  \label{prvo2}
\end{eqnarray}
In all these expressions for the comoving volume $V(t,y^\mu)$, $M$ is
a time-independent scalar related to the energy density through equation~(\ref{li7}):
\begin{eqnarray}
\rho = \frac{M}{V} \,,
\end{eqnarray}
and a first integral of the energy-momentum conservation equation
$\dot{\rho}+\Theta\rho =0$.  
Taking into account that the proper
volume~(\ref{prvo}) is defined up to a multiplicative time-independent scalar
[see Eq.~(\ref{solu})], we naturally choose $M$ to be a positive constant, and
in fact we could even choose its value.  On the other hand, the proper volume is
related to the local scale factor $a(\tau,y^\mu)$ by the expression
\[ \frac{\dot{V}}{V}= 3\frac{\dot{a}}{a}\equiv 3H\,, \hspace{12mm}
a^2\equiv [\mbox{det}(h_{\alpha\beta})]^{\textstyle{1\over3}}\,, \]
where $H$ is the local Hubble parameter.

Since we will be mainly interested in the behaviour near the singularities
we can neglect the effect of the cosmological constant and consider only
the case (iii) given by equations~(\ref{prvo},\ref{prvo2}).  This well-known
fact can be checked from the expressions we obtain for $V(t,y^\mu)$, from
where one can see that in the limit $\tau\rightarrow{\cal T}$ all the cases behave
like the case (iii), i.e., $V\sim \tau-{\cal T}$.
In any case, the procedure we will follow here to solve the different equations
can be applied from the beginning to the end to all the cases with a non-zero
cosmological constant.

Then, as we can see from~(\ref{prvo},\ref{prvo2}), and assuming without loss of
generality $\Sigma\geq 0$ ($\Rightarrow \delta{\cal T}>0$), for
$\tau > {\cal T}$ we have an expanding fluid (note that this depends on which
fluid element we are looking at, i.e. on $y^\mu$), whereas for $\tau < {\cal T}
- \delta{\cal T}$ we have a collapsing one.  The region $\tau\in
({\cal T}-\delta{\cal T},{\cal T})$ is excluded by the requirement of
having a positive energy density.  Then, for an expanding IDM, ${\cal T}(y^\mu)$
is the Bang time corresponding to a fluid element with comoving
coordinates $y^\mu$. In the same way, ${\cal T}-\delta{\cal T}$ is
the Crunch time for a contracting IDM.

The next step is to solve the equations for the triad vectors
$\{\mb{e}_\alpha\}$ [Eq.~(\ref{base})], which become
\[ \dot{e}_1{}^\mu = -\textstyle{1\over3}(\Theta-2\sigma_+)e_1{}^\mu\,,\]
\[ \dot{e}_2{}^\mu = -\textstyle{1\over3}(\Theta+\sigma_++
\textstyle{\sqrt{3}}\sigma_-)e_2{}^\mu\,,\]
\[ \dot{e}_3{}^\mu = -\textstyle{1\over3}(\Theta+\sigma_+-
\textstyle{\sqrt{3}}\sigma_-)e_3{}^\mu\,.\]
The result, in the case $\Lambda=0\,,$ is
\begin{equation}
e_\alpha{}^\mu = b_\alpha{}^\mu(\tau-{\cal T})^{-p_\alpha}
(\tau-{\cal T}+\delta{\cal T})^{-q_\alpha}\,,\label{tria}
\end{equation}
where the quantities $b_\alpha{}^\mu$ in (\ref{tria}) are
the time-independent components of a triad defining a Riemannian
3-dimensional geometry whose metric tensor is given by
$q_{\mu\nu}=\delta_{\alpha\beta}
b^{\alpha}{}_\mu b^{\beta}{}_\nu$, being
$b^{\alpha}{}_\mu$ the inverse matrix of $b_\alpha{}^\mu$.
Moreover, $q_\alpha\equiv 2/3-p_\alpha$ and
\begin{eqnarray}
\fl p_1\equiv\frac{\sqrt{3}\Sigma-2\Sigma_+}
{3\sqrt{3}\Sigma}\,,\hspace{5mm}
p_2\equiv\frac{\Sigma_++\sqrt{3}(\Sigma+\Sigma_-)}
{3\sqrt{3}\Sigma}\,,\hspace{5mm}
p_3\equiv\frac{\Sigma_++\sqrt{3}(\Sigma-\Sigma_-)}
{3\sqrt{3}\Sigma}\,,\label{tikasner}
\end{eqnarray}
are defined in the analogous way as they are defined for Bianchi I
IDMs.  Indeed, we can check that $p_\alpha$ and $q_\alpha$ satisfy
the Kasner relations
\[\sum_{\alpha=1}^3 p_\alpha = \sum_{\alpha=1}^3 p^2_\alpha = 1
 ~ \Longleftrightarrow ~
\sum_{\alpha=1}^3 q_\alpha = \sum_{\alpha=1}^3 q^2_\alpha = 1 \,, \]
so we will call them {\em local Kasner coefficients}.  It is convenient,
for practical purposes,  to consider the fact that one can
assume, without loss of generality, that in an open domain of a given
fluid element located at $y^\mu$, the Kasner coefficients can be
chosen to lie in the following intervals:
\begin{equation}
p_1\in\left(-\textstyle{1\over3},0\right)\,,~~~~
p_2\in\left(0,\textstyle{2\over3}\right)\,,~~~~
p_3\in\left(\textstyle{2\over3},1\right)\,. \label{kasner}
\end{equation}
Then, looking at (\ref{tria}) it is clear that the first-order
solution has an initial singularity (in general not simultaneous)
of the Kasner type, also known as a velocity-dominated singularity~\cite{ELS}.

Summarizing, expressions (\ref{solu}), (\ref{prvo},\ref{prvo2}),
and (\ref{tria}) solve the
first-order evolution.  From them we can find the time dependence
of all the variables.   With regard to the spatial dependence,
which is encoded inside the functions ${\cal T}$, $\{\mb{b}_\alpha\}$,
and $\Sigma_\pm$, the only restriction
comes from the constraint~(\ref{monc}) (the {\em momentum constraint})
which is equivalent to the following three equations
\begin{equation}
\mb{\partial}_1(\sigma_++\Theta)-3a_1\sigma_+-\sqrt{3}n_{23}
\sigma_-=0\,, \label{moc1}
\end{equation}
\begin{equation}
\mb{\partial}_2(\sigma_++\sqrt{3}\sigma_--2\Theta)-3a_2(\sigma_+
+\sqrt{3}\sigma_-)+\sqrt{3}n_{13}(\sqrt{3}\sigma_+-\sigma_-)=0\,,
\label{moc2}
\end{equation}
\begin{equation}
\mb{\partial}_3(\sigma_+-\sqrt{3}\sigma_--2\Theta)-3a_3(\sigma_+
-\sqrt{3}\sigma_-)-\sqrt{3}n_{12}(\sqrt{3}\sigma_++\sigma_-)=0\,,
\label{moc3}
\end{equation}
To see which kind of restrictions these equations impose we need to
compute the quantities $a_\alpha$ and $n_{\alpha\beta}$, or equivalently
the commutators $\gamma^\alpha{}_{\beta\lambda}$.  Using the
expressions~(\ref{spco}) we get
\begin{equation}
\fl \gamma^\alpha{}^{}_{\beta\lambda} = \left\{
\gamma'^{\underline{\alpha}}{}^{}_{\underline{\beta\lambda}}
+2\delta^{\underline{\alpha}}{}^{}_{\![\underline{\beta}} \left[
(\mb{\partial'}^{}_{\underline{\lambda}]} p^{}_{\underline{\alpha}})
\ln \frac{P}{Q}
+ \frac{\mb{\partial'}^{}_{\underline{\lambda}]} P}{P}
p_{\underline{\alpha}}
+ \frac{\mb{\partial'}^{}_{\underline{\lambda}]} Q}{Q}
q_{\underline{\alpha}} \right]\right\}
P^{p^{}_{\underline{\lambda}}-p^{}_{\underline{\alpha}}
-p^{}_{\underline{\beta}}}\;Q^{q^{}_{\underline{\lambda}}-
q^{}_{\underline{\alpha}}-q^{}_{\underline{\beta}}}\,, \label{comf}
\end{equation}
where $\gamma'^\alpha_{\beta\delta}$ are the (time-independent)
commutator functions associated with the triad $\{\mb{b}_\alpha\}$,
underlined indices do not follow the usual index summation convention,
and we have used the following definitions:
\[ \mb{\partial'}_\alpha\equiv b_\alpha{}^\mu\partial_{y^\mu}
\,, \hspace{4mm}
P \equiv \tau-{\cal T}\,, \hspace{4mm}
Q \equiv \tau-{\cal T}+\delta{\cal T}\,. \]
Then, from~(\ref{coco},\ref{comf}) we have
\begin{eqnarray}
\fl a_\alpha = \left\{a'_{\underline{\alpha}} +\frac{1}{2}\left[
(\mb{\partial'}^{}_{\underline{\alpha}} p^{}_{\underline{\alpha}})
\ln \frac{P}{Q} + (p_{\underline{\alpha}}-1)
\frac{\mb{\partial'}^{}_{\underline{\alpha}} P}{P}
+ (q_{\underline{\alpha}}-1)
\frac{\mb{\partial'}^{}_{\underline{\alpha}} Q}{Q}\right]\right\}
P^{-p^{}_{\underline{\alpha}}}\;Q^{-q^{}_{\underline{\alpha}}}
\nonumber   \,,
\end{eqnarray}
\begin{eqnarray}
\fl n^{\alpha\beta} = \left\{ \begin{array}{ll}
n'^{\underline{\alpha\alpha}}\;P^{2p^{}_{\underline{\alpha}}-1}
\;Q^{2q^{}_{\underline{\alpha}}-1} & \mbox{if $\alpha=\beta$\,,} \\
\mbox{} & \mbox{} \\
\left\{ n'^{\underline{\alpha\beta}}-\frac{1}{2}
\epsilon^{\underline{\alpha\beta\delta}}
\mb{\partial'}^{}_{\underline{\delta}}\left[
(p_{\underline{\alpha}}-p_{\underline{\beta}})\ln  \frac{P}{Q}
\right]\right\}
P^{-p^{}_{\underline{\delta}}}\;Q^{-q^{}_{\underline{\delta}}}
~~ (\alpha\neq\delta\neq\beta)
& \mbox{if $\alpha\neq\beta$\,.}
\end{array} \right. \nonumber
\end{eqnarray}
Introducing all these expressions into the constraints
(\ref{moc1}-\ref{moc3}), we can see that they are equivalent to
the following three time-independent equations
\begin{equation}
\fl \mb{\partial'}_1\left[\Sigma_++\textstyle{3\over2}M
({\cal T}-\textstyle{1\over2}\delta{\cal T})\right] =
3 a'_1\Sigma_+ + \sqrt{3}n'_{23}\Sigma_-\,,
\label{cos1}
\end{equation}
\begin{equation}
\fl \mb{\partial'}_2\left[\Sigma_++\sqrt{3}\Sigma_-
-3M({\cal T}-\textstyle{1\over2}\delta{\cal T})\right] =
3a'_2(\Sigma_++\sqrt{3}\Sigma_-) -
\sqrt{3}n'_{13}(\sqrt{3}\Sigma_+-\Sigma_-)\,, \label{cos2}
\end{equation}
\begin{equation}
\fl \mb{\partial'}_3\left[\Sigma_+-\sqrt{3}\Sigma_-
-3M({\cal T}-\textstyle{1\over2}\delta{\cal T})\right] =
3a'_3(\Sigma_+-\sqrt{3}\Sigma_-) +
\sqrt{3}n'_{12}(\sqrt{3}\Sigma_++\Sigma_-)\,. \label{cos3}
\end{equation}
This is not an obvious result since equations~(\ref{moc1})-(\ref{moc3})
are linear combinations of functions depending on $\{\tau,y^\mu\}$
with coefficients depending only on $\{y^\mu\}$ and, in principle, one
would expect that from each of them we would get several equations.
What we have seen is that we only get one for each of them.
Equations (\ref{cos1})-(\ref{cos3}) are then constraints on the quantities
${\cal T}$, $\{\mb{b}_\alpha\}$ and $\Sigma_\pm\,.$

At this point, the construction of the first-order solution is
essentially finished.  As we have seen, the time dependence is
explicitly known whereas the spatial one is encoded in the
quantities ${\cal T}$, $\{\mb{b}_\alpha\}$ and $\Sigma_\pm\,.$
Now, we can split the zero and first order terms by taking
into account that the zero order is characterized by the absence
of any spatial gradient.  Then, the splitting is given by the
relations
\begin{equation}
b_\alpha{}^\mu = \delta_\alpha{}^\mu +
   {}_1b_\alpha{}^\mu \,, \label{spte}
\end{equation}
\begin{equation}
\Sigma_\pm = {}_o\Sigma_\pm + {}_1\Sigma_\pm \,, \hspace{4mm}
{\cal T} =  {}_o{\cal T} + {}_1{\cal T} \,, \label{spl2}
\end{equation}
where ${}_o\Sigma_\pm$ and ${}_o{\cal T}$ are constants.  From~(\ref{spte})
we deduce that the commutator functions $\gamma'^\alpha_{\beta\delta}$,
or equivalently, $a'_\alpha$ and $n'^{\alpha\beta}$ do not have a
zero-order part.  Moreover, it also gives a splitting
of the spatial derivative $\mb{\partial'}_\alpha$
\begin{equation}
\mb{\partial'}_\alpha = \partial_{y^\mu} +
{}_1b_\alpha{}^\mu\partial_{y^\mu} \,. \label{spl1}
\end{equation}
Then, it also follows that equations (\ref{cos1})-(\ref{cos3}) are automatically
satisfied at zero-order.  At first order they are
\begin{eqnarray}
\fl \partial_1\left[ {}_1\Sigma_+ + \Xi \right] =
3a'_1{}_o\Sigma_+ + \sqrt{3}n'_{23}{}_o\Sigma_-\,, \nonumber
\end{eqnarray}
\begin{eqnarray}
\fl \partial_2\left[ {}_1\Sigma_+ + \sqrt{3}{}_1\Sigma_- - 2\Xi \right] =
3a'_2({}_o\Sigma_++\sqrt{3}{}_o\Sigma_-) - \sqrt{3}n'_{13}
(\sqrt{3}{}_o\Sigma_+-{}_o\Sigma_-)\,, \nonumber
\end{eqnarray}
\begin{eqnarray}
\fl \partial_3\left[ {}_1\Sigma_+ - \sqrt{3}{}_1\Sigma_- - 2\Xi \right] =
3a'_3({}_o\Sigma_+-\sqrt{3}{}_o\Sigma_-) + \sqrt{3}n'_{12}
(\sqrt{3}{}_o\Sigma_++{}_o\Sigma_-)\,, \nonumber
\end{eqnarray}
where
\[ \Xi\equiv \frac{3}{2}M{}_1{\cal T} -
\frac{1}{\sqrt{3}{}_o\Sigma}({}_o\Sigma_+{}_1\Sigma_++{}_o\Sigma_-
{}_1\Sigma_-)\,, \hspace{4mm}
{}_o\Sigma^2\equiv\textstyle{1\over3}({}_o\Sigma^2_++{}_o\Sigma^2_-) \,. \]
It is worth noting that our zero order solution is exactly Bianchi I,
and so is the first order part from the point of view of the evolution
equations.

To complete the construction of the first-order solution, we give the
expression for the remaining quantities, that is, the gravito-electric and
-magnetic tensor fields
$E_{\alpha\beta}$ and $H_{\alpha\beta}\,.$
As we said before, equation~(\ref{gete}) tells us that $E_{\alpha\beta}$
is diagonal since $\sigma_{\alpha\beta}$ is diagonal.  Then, we only
need to compute the quantities $E_+$ and $E_-$ defined as in~(\ref{depm}).
The result is
\[ E_+ = \textstyle{1\over3}\Theta\sigma_+ +\textstyle{1\over3}(
\sigma^2_+ - \sigma^2_-) = \frac{1}{3V^2}\left\{ \frac{3}{2}M(\tau-{\cal T}+
\frac{1}{2}\delta{\cal T})\Sigma_+ + \Sigma^2_+ -
\Sigma^2_- \right\} \,, \]
\[ E_- = \textstyle{1\over3}(\Theta-2\sigma_+)\sigma_-=
\frac{1}{3V^2}\left\{ \frac{3}{2}M(\tau-{\cal T}+
\frac{1}{2}\delta{\cal T})-2\Sigma_+\right\}\Sigma_-\,. \]
The gravito-magnetic field $H_{\alpha\beta}$ has five independent
components, which can be given in terms of the quantities
$H_\pm$ [defined as in Eq.~(\ref{depm})], and $H_1\equiv\sqrt{3}H_{23}\,,$
$H_2\equiv\sqrt{3}H_{13}\,,$ and $H_3\equiv\sqrt{3}H_{12}\,.$
Their form is given by
\begin{eqnarray}
\fl H_+ & = & -\textstyle{3\over2}n_{11}\sigma_+ -\textstyle{\sqrt{3}\over2}
(n_{22}-n_{33})\sigma_- \nonumber \\
\fl & = & -\frac{2\Sigma_+}{M}n'_{11}P^{2(p_1-1)}
Q^{2(q_1-1)}-\frac{2\Sigma_-}{\sqrt{3}M}\left\{n'_{22}
P^{2(p_2-1)}Q^{2(q_2-1)}-n'_{33}P^{2(p_3-1)}Q^{2(q_3-1)} \right\} \,, \nonumber
\end{eqnarray}
\begin{eqnarray}
\fl H_-& = & -\textstyle{\sqrt{3}\over2}(n_{22}-n_{33})\sigma_+
+\textstyle{1\over2}(n_{11}-2n_{22}-2n_{33})\sigma_- \nonumber \\
\fl & = &-\frac{2\Sigma_+}{\sqrt{3}M}\left\{ n'_{22}
P^{2(p_2-1)}Q^{2(q_2-1)}-n'_{33}P^{2(p_3-1)}Q^{2(q_3-1)}\right\}
\nonumber \\
\fl & & +\frac{2\Sigma_-}{3M}\left\{n'_{11}P^{2(p_1-1)}
Q^{2(q_1-1)}-2n'_{22}P^{2(p_2-1)}Q^{2(q_2-1)}-2n'_{33}
P^{2(p_3-1)}Q^{2(q_3-1)}\right\} \,, \nonumber
\end{eqnarray}
\begin{eqnarray}
\fl H_1 & = & -\sqrt{3}n_{23}\sigma_+ + (\mb{\partial}_1-a_1)\sigma_-
\nonumber \\
\fl & = & \textstyle{4\over{3M}}\left\{  \partial_1({}_1\Sigma_-) -
{}_o\Sigma_-a'_1 - \sqrt{3}{}_o\Sigma_+n'_{23}
 -{}_o\Sigma_-\frac{\partial_1 V}{V} +
\frac{\sqrt{3}}{2}{}_o\Sigma_+\partial_1\left[(p_2-p_3)\ln\frac{P}{Q}
\right] \right. \nonumber \\
\fl & & \left. -\textstyle{1\over2}{}_o\Sigma_-\left[(\partial_1 p_1)
\ln\frac{P}{Q}+
(p_1-1)\frac{\partial_1 P}{P}+ (q_1-1)\frac{\partial_1 Q}{Q}\right]\right\}
P^{-p_1-1}Q^{-q_1-1} \,, \nonumber
\end{eqnarray}
\begin{eqnarray}
\fl H_2 & = &\textstyle{3\over2}n_{13}(\sigma_+ + \sqrt{3}\sigma_-)
+\textstyle{1\over2}(\mb{\partial}_2-a_2)(\sqrt{3}\sigma_+ -\sigma_-)
\nonumber \\
\fl & = & \textstyle{2\over{3M}}\left\{ \partial_2(\sqrt{3}{}_1\Sigma_+-
{}_1\Sigma_-)-(\sqrt{3}{}_o\Sigma_+-{}_o\Sigma_-)a'_2
+3({}_o\Sigma_++\sqrt{3}{}_o\Sigma_-)n'_{13}\right. \nonumber \\
\fl & & -(\sqrt{3}{}_o\Sigma_+-{}_o\Sigma_-)\textstyle{{\partial_2 V}\over{V}}
+\frac{3}{2}({}_o\Sigma_++\sqrt{3}{}_o\Sigma_-)\partial_2\left[(p_1-p_3)
\ln\frac{P}{Q}\right]\nonumber \\
\fl & & \left. -\textstyle{1\over2}(\sqrt{3}{}_o\Sigma_+-{}_o\Sigma_-)
\left[(\partial_2 p_2)\ln\frac{P}{Q}+
(p_2-1)\frac{\partial_2 P}{P}+(q_2-1)\frac{\partial_2 Q}{Q}\right]\right\}
P^{-p_2-1}Q^{-q_2-1} \,, \nonumber
\end{eqnarray}
\begin{eqnarray}
\fl H_3 & = & \textstyle{3\over2}n_{12}(\sigma_+ - \sqrt{3}\sigma_-)
-\textstyle{1\over2}(\mb{\partial}_3-a_3)(\sqrt{3}\sigma_+ +\sigma_-)
\nonumber \\
\fl & = & \textstyle{2\over{3M}}\left\{ -\partial_3(\sqrt{3}{}_1\Sigma_++
{}_1\Sigma_-)+(\sqrt{3}{}_o\Sigma_++{}_o\Sigma_-)a'_3
+3({}_o\Sigma_+-\sqrt{3}{}_o\Sigma_-)n'_{12}\right. \nonumber \\
\fl & & +(\sqrt{3}{}_o\Sigma_++{}_o\Sigma_-)\textstyle{{\partial_3 V}\over{V}}
-\frac{3}{2}({}_o\Sigma_+-\sqrt{3}{}_o\Sigma_-)\partial_3\left[(p_1-p_2)
\ln\frac{P}{Q}\right]\nonumber \\
\fl & & \left. +\textstyle{1\over2}(\sqrt{3}{}_o\Sigma_++{}_o\Sigma_-)
\left[(\partial_3 p_3)\ln\frac{P}{Q}+
(p_3-1)\frac{\partial_3 P}{P}+(q_3-1)\frac{\partial_3 Q}{Q}\right]\right\}
P^{-p_3-1}Q^{-q_3-1} \,. \nonumber
\end{eqnarray}

In order to understand the dynamical behaviour of IDMs using the first-order
approximation of the LWAS, we have computed the asymptotic behaviour in time,
both near the Big-Bang singularity ($\tau\rightarrow{\cal T}$) and for large
future proper time ($\tau\rightarrow\infty$) of the main physical quantities
that are used in the covariant fluid approach.
The results are summarized in Table~\ref{tasym}.  These results represent
the generic behaviour of IDMs, there are particular cases, characterized 
by the vanishing of some of the connection quantities $n_{\alpha\beta}$ or 
some the components of the gradient of ${\cal T}$, in which the behaviour 
indicated in Table~\ref{tasym} has to be revised.  For the sake of brevity 
we have not included these particular cases here.
Moreover, to obtain this asymptotic behaviour we have used the fact that
in the neighbourhood of a fluid element one can always assume that the Kasner
coefficients are distributed as in~(\ref{kasner}).

\begin{table}
\caption{Asymptotic behaviour of the first-order solution.\label{tasym}}
\begin{indented}
\item[]\begin{tabular}{@{}cccc}
\br
\mbox{Quantity}  & \mbox{Behaviour as~} $\tau\rightarrow{\cal T}$ &
\mbox{Behaviour as~} $\tau\rightarrow\infty$ \\
\mr
$\Theta$ \parbox[c][8mm]{1mm}{\mbox{}}  & $(\tau-{\cal T})^{-1}$ & $2\tau^{-1}$
\\
$\sigma_\pm$  \parbox[c][8mm]{1mm}{\mbox{}}  &
$\frac{4\Sigma_\pm}{3M\delta{\cal T}}(\tau-{\cal T})^{-1}$
& $\frac{4\Sigma_\pm}{3M}\tau^{-2}$
\\
$\frac{\sigma}{\Theta}$  \parbox[c][8mm]{1mm}{\mbox{}}  &
$\frac{4\Sigma}{3M\delta{\cal T}}$  &
$\frac{2\Sigma}{3M}\tau^{-1}$
\\
$\frac{\rho}{\Theta^2}$  \parbox[c][8mm]{1mm}{\mbox{}}  &
$\frac{4\Sigma_\pm}{3M\delta{\cal T}}(\tau-{\cal T})$  &
$\frac{1}{3}$
\\
$\frac{E_+}{\Theta^2}$  \parbox[c][1cm]{1mm}{\mbox{}}  &
$\textstyle{1\over3}\left\{\Sigma_++\frac{4(\Sigma^2_+-
\Sigma^2_-)}{3M\delta{\tau}}\right\}\frac{4}{3M\delta{\cal T}}$  &
$\frac{2\Sigma_+}{9M}\tau^{-1}$
\\
$\frac{E_-}{\Theta^2}$  \parbox[c][1cm]{1mm}{\mbox{}}  &
$\textstyle{1\over3}\left\{1-\frac{8\Sigma_+}
{3M\delta{\tau}}\right\}\frac{4\Sigma_-}{3M\delta{\cal T}}$  &
$\frac{2\Sigma_-}{9M}\tau^{-1}$
\\
$\frac{H_+}{\Theta^2}$ \parbox[c][8mm]{1mm}{\mbox{}}  &
${\cal H}^{BB}_+(y^\mu)(\tau-{\cal T})^{2p_1}$  &
${\cal H}^{\infty}_+(y^\mu)\tau^{-2/3}$
\\
$\frac{H_-}{\Theta^2}$  \parbox[c][8mm]{1mm}{\mbox{}}  &  
${\cal H}^{BB}_-(y^\mu)(\tau-{\cal T})^{2p_1}$   &
${\cal H}^{\infty}_-(y^\mu)\tau^{-2/3}$
\\
$\frac{H_1}{\Theta^2}$  \parbox[c][8mm]{1mm}{\mbox{}}  &
${\cal H}^{BB}_1(y^\mu)(\tau-{\cal T})^{-p_1}$   &
${\cal H}^{\infty}_1(y^\mu)\tau^{-2/3}$
\\
$\frac{H_2}{\Theta^2}$  \parbox[c][8mm]{1mm}{\mbox{}}  &
${\cal H}^{BB}_2(y^\mu)(\tau-{\cal T})^{-p_2}$   &
${\cal H}^{\infty}_2(y^\mu)\tau^{-2/3}$
\\
$\frac{H_3}{\Theta^2}$ \parbox[c][8mm]{1mm}{\mbox{}}   &
${\cal H}^{BB}_3(y^\mu)(\tau-{\cal T})^{-p_3}$   &
${\cal H}^{\infty}_3(y^\mu)\tau^{-2/3}$
\\
\br
\end{tabular}
\end{indented}
\end{table}


\section{On the validity of the Long Wavelength approximation scheme}
\label{sec4}

The LWAS is based on the general idea that spatial gradients can
be neglected with respect to time derivatives.  This assumption
puts restrictions to the validity of the expressions we have
given above.  This issue was already studied by
Deruelle and Langlois in~\cite{LADR}. Their criterion to consider
the approximation as valid is to compare the third-order terms
arising from the first-order solution and to check whether or
not they are smaller than the first-order ones.  To that end,
they computed the neglected terms from the first-order solution,
essentially the spatial curvature terms ${}^3R$ and ${}^3S_{\mu\nu}$.
Then, they realized that in general the effect of the local anisotropy
makes some of spatial curvature components to blow up.  Then,
they argue that when this happens one shall recover the
oscillatory BKL behaviour.

Our calculations agree with their results.  First, for late times
($\tau\rightarrow\infty$), the system evolves towards an Einstein-de Sitter
model as expected (see Table~\ref{tasym}), whereas for $\Lambda > 0$ it
would evolve toward the de~$\!$Sitter model.   On the contrary,  we have found
that indeed some components of the spatial curvature blow up in general as one
approaches the initial singularity ($\tau-{\cal T}\rightarrow 0$).
Nevertheless, here we want to take a complementary point of view to deal with
this issue.   In particular, we only need to analyze the quantities used in the
covariant fluid approach constructed the first-order solution.
In our scheme there is a set of main variables, namely
$\{e_{\alpha}{}^\mu,\Theta,\sigma_{\alpha\beta}\}$, and
a set of auxiliary variables, $\{\gamma^\delta_{\alpha\beta},\rho,
E_{\alpha\beta},H_{\alpha\beta}\}$.  The idea being that the secondary
variables can be computed from the main ones.  Of particular interest
are the gravito-electric and -magnetic fields.  Their first-order
form has been given in Table~\ref{tasym}.  As we can see, their
behaviour near the initial singularity is completely different.
While the normalized\footnote{Note that we are normalizing using the
expansion, which is related to the Hubble length, the natural scale
of the system, by $\Theta=3H\,.$} gravito-electric field,
$\Theta^{-2}E_{\alpha\beta}$,
a dimensionless quantity, is finite there (but depending on the fluid
element we consider), the normalized gravito-magnetic field,
$\Theta^{-2}H_{\alpha\beta}$, has components that blow up.
These divergences constitute an indication of  the breakdown of
the approximation scheme as we approach the initial singularity.
Here again it is the local anisotropy that produces the blowing up
of the gravito-magnetic field: this first-order quantity 
becomes dominant over the zero-order ones and the scheme breaks down.
This can be seen in Table~\ref{tasym}, where the dependence of the
components of $\Theta^{-2}H_{\alpha\beta}$ on the Kasner coefficients
$p_\alpha$ is explicit.  In contrast, the gravito-electric field
does not depend on them and its normalization is regular.
Moreover, from Table~\ref{tasym} we also see the fact that the
effect of matter, described by the normalized energy density
$\Theta^{-2}\rho$ (essentially, the cosmological density parameter
$\Omega$), can be neglected near the initial singularity, a
well-known result.

Another interesting difference between the behaviour of the
gravito-electric and -magnetic tensors is their local/non-local
character.  Whereas $E_{\alpha\beta}$ is completely local at
first-order, that is, for a fixed fluid element $y^\alpha$ it only
depends on the value of the main variables at that point [see
Eq.~(\ref{gete})], $H_{\alpha\beta}$ does depend on spatial
gradients of the main quantities.  More specifically, on
the shear-rate spatial gradients [see Equation~(\ref{magw})].
In this sense, and in the spirit of the LWAS, $H_{ab}$ is a
purely first-order quantity and the spatial Ricci tensor is a
purely second-order quantity.   The blowing up of $H_{\alpha\beta}$
(at first order) and ${}^3R_{\alpha\beta}$ (as computed from the
first-order solution) are not independent facts.  To
understand this we have to look at the evolution of the spatial
Ricci tensor:
\begin{eqnarray}
{}^3\dot{R} = -\textstyle{2\over3}\Theta\;{}^3R -
2\sigma^{\alpha\beta}\;{}^3S_{\alpha\beta}\,, \label{ricci1}
\end{eqnarray}
\begin{eqnarray}
{}^3\dot{S}_{\alpha\beta} = \textstyle{2\over3}\Theta\;{}^3S_{\alpha\beta}
+\sigma_{<\alpha}{}^\delta\;{}^3S_{\beta>\delta}-\textstyle{1\over6}
\sigma_{\alpha\beta}\;{}^3R + \mbox{curl}(H)_{\alpha\beta}\,.
\label{ricci2}
\end{eqnarray}
In IDMs  the gravito-magnetic field has its origin in the anisotropy,
$H_{ab}=\mbox{curl}(\sigma)_{ab}$, and it is the only source for the Ricci
tensor as we can see from equations~(\ref{ricci1},\ref{ricci2}).   
First, it can generate ${}^3S_{ab}$ through equation~(\ref{ricci2}),
and ${}^3S_{ab}$ is a source for the scalar spatial curvature
${}^3R$ in equation~(\ref{ricci1}).

On the other hand, the dynamical picture we have when approaching
the spacelike singularity is essentially the BKL picture,
in which the system has a Kasner-like evolution with some
determined Kasner coefficients, say $p'_\alpha$, and at some
point the system jumps to another Kasner phase characterized
by other Kasner coefficients, say $p''_\alpha$.
Within our approach, we can define the following time-dependent
Kasner coefficients
\begin{eqnarray}
\fl p_1\equiv\frac{\sqrt{3}\sigma-2\sigma_+}
{3\sqrt{3}\sigma}\,,\hspace{5mm}
p_2\equiv\frac{\sigma_++\sqrt{3}(\sigma+\sigma_-)}
{3\sqrt{3}\sigma}\,,\hspace{5mm}
p_3\equiv\frac{\sigma_++\sqrt{3}(\sigma-\sigma_-)}
{3\sqrt{3}\sigma}\,.\label{tdkasner}
\end{eqnarray}
For the first-order solution, these become the time-independent ones 
defined in~(\ref{tikasner}), because all the
components of the shear have the same time dependence.  
To see how they change at higher
order within the LWAS, it is enough to consider the
evolution equation for the shear~(\ref{shea}) with the trace-free
part of the three-dimensional Ricci tensor, ${}^3S_{\alpha\beta}$
(which is a second-order quantity) computed from the
first-order solution.   Now, in order to write down the
evolution equations for the shear it is important to consider
some important facts.  First, that we are considering a basis
in which the shear tensor diagonalizes, so it can be described
only by the scalars $(\sigma_+,\sigma_-)$.  Second,
in the computation of ${}^3S_{\alpha\beta}$ one can see
that the diagonal terms dominate over the non-diagonal
ones as one approaches the initial singularity at 
$\tau\rightarrow{\cal T}$.  Therefore,
for our purposes we can consider that ${}^3S_{\alpha\beta}$
is diagonal and described by the two quantities
\begin{eqnarray}
{}^3S_+\equiv -\textstyle{3\over2}\;{}^3S_{11}\,, \hspace{4mm}
{}^3S_-\equiv \textstyle{\sqrt{3}\over2}(\;{}^3S_{22}-\;{}^3S_{33})\,,
\nonumber 
\end{eqnarray}
This also means that within this approximation the shear and the
trace-free Ricci tensor commute, which in turn implies that the
gravitomagnetic field is transverse: $\mbox{div}(H)_a =0$.   
Taking into account all these considerations, the evolution equations 
for the Kasner coefficients~(\ref{tdkasner}) can be expressed 
in a compact way by introducing the following vector: 
$\vec{\mbox{p}} = (p_1,p_2,p_3)$.  And the evolution of
$\vec{\mbox{p}}$ is given by
\begin{eqnarray}
\dot{\vec{\mbox{p}}} = \Omega\,{\cal A}\,\vec{\mbox{p}}\,,
\label{pdot}
\end{eqnarray}
where
\begin{eqnarray}
\Omega\equiv \frac{\sigma_+ \; {}^3S_- - \sigma_- \; {}^3S_+}
{3\sqrt{3}\sigma^2}\,,~~~~
{\cal A}\equiv \left(\begin{array}{ccc}
0 & -1 & 1 \\
1 & 0  & -1 \\
-1 & 1  & 0 \end{array}\right) \,, \nonumber
\end{eqnarray}
where ${}^3S_\pm+$ are computed from the first-order solution, and
$\sigma_\pm$ from the zero-order one.
Equation (\ref{pdot}) can be solved for each fluid element and the
solution can expressed in the following form
\begin{eqnarray}
\fl \vec{\mbox{p}}(\tau,y^\mu) = \vec{\mbox{p}}_0(y^\mu) +
\frac{1-\cos\left[\sqrt{3}\,I(\tau,y^\mu)\right]}{3}\,
{\cal A}^2\vec{\mbox{p}}_0(y^\mu)+\frac{1}{\sqrt{3}}
\sin\left[\sqrt{3}\,I(\tau,y^\mu)\right]\,{\cal A}
\vec{\mbox{p}}_0(y^\mu)\,, \label{cambio}
\end{eqnarray}
where
\begin{eqnarray}
I(\tau,y^\mu)\equiv\int_{\tau_0}^\tau \Omega(\tau',y^\mu)d\tau'\,,
~~~~ {\cal A}^2 = \left(\begin{array}{ccc}
-2 &  1 & 1 \\
 1 & -2 & 1 \\
 1 &  1 & -2 \end{array} \right)    \,. \nonumber
\nonumber
\end{eqnarray}
From the first order solution we find that in the generic case,
i.e. without extra assumptions, the quantity $I(\tau, y^\mu)$
behaves, near the initial singularity, as
\begin{eqnarray}
I(\tau,y^\mu)~~\stackrel{\tau\rightarrow{\cal T}}{\longrightarrow}
~~{\cal I}(y^\mu)(p_1-p_2)(\tau-{\cal T})^{-2p_3} \,. \nonumber
\end{eqnarray}
Then, equation~(\ref{cambio}) shows the change that the Kasner
coefficients will suffer when the spatial curvature is taking
into account.  However, this equation will not be valid once
the Kasner coefficient have changed their value significally,
since we are approximating the spatial curvature from the
first-order solution, where the Kasner coefficients are {\em static}.
This means that the behaviour of the spatial curvature along
the different spatial directions does not change.  In order 
to take into account this fact one should compute the spatial
curvature from equations~(\ref{ricci1},\ref{ricci2}), which 
describe the evolution of ${}^3R$ and ${}^3S_{\alpha\beta}$,
mainly driven by a second-order quantity in the LWAS:
$\mbox{curl}(H)_{ab}$.   This in agreement with the observation
in numerical simulations of the homogeneous Bianchi IX models,
which exhibit a BKL behaviour, that the gravito-magnetic
tensor $H_{ab}$ generates the transition between different
Kasner epochs (see~\cite{WAEL}).

Finally, we want to remark that the discussion presented here
is based on the generic solution of the LWAS.  One can restrict
the free functions and parameters that appear in the first-order
solution in order to make it well behaved near the initial
singularity.   More specifically, we can impose the following
restrictions on the first-order solution:
\begin{eqnarray}
{\cal T} = {\cal T}_0:~\mbox{constant}\,,~~~\mbox{and}~~~n'_{11}=0\,.
\label{parti}
\end{eqnarray}
The first condition means that the initial singularity must be
simultaneous.  Actually, without loss of generality we can choose
${\cal T}_0 = 0$. The second condition in~(\ref{parti}) is a
restriction on the triad $\{\mb{b}_\alpha\}$.  Then, under
conditions~(\ref{parti}) one can check that
\begin{eqnarray}
\frac{H_{\alpha\beta}}{\Theta^2}~~
\stackrel{\tau\rightarrow{\cal T}}{\longrightarrow}~~
0\,,~~~~~ \frac{{}^3R_{\alpha\beta}}{\Theta^2}~~
\stackrel{\tau\rightarrow{\cal T}}{\longrightarrow}~~
0\,. \label{regular}
\end{eqnarray}
With respect to the results shown in Table~\ref{tasym} the only
thing that changes is the behaviour of $H_{ab}$ near the
singularity:  the precise behaviour can be found in Table~\ref{tabla2}.
Although for the component $\Theta^{-2}H_3$ the expression in
Table~\ref{tabla2} suggests a logarithmic singularity for the
case $p_3=1\,,~p_1=p_2=0$, this is not the case.  For
this very particular case, the coefficient ${\cal H}^{0}_3(y^\mu)$
vanishes identically for that situation and hence the
result~(\ref{regular}) is general for the particular
case~(\ref{parti}).

\begin{table}
\caption{Asymptotic behaviour of the first-order solution in the particular
case characterized by the relations~(\ref{parti}).\label{tabla2}}
\begin{indented}
\item[]\begin{tabular}{@{}ccc}
\br
\mbox{Quantity}  & \mbox{Behaviour as~} $\tau\rightarrow 0$  \\
\mr
$\frac{H_+}{\Theta^2}$ \parbox[c][8mm]{1mm}{\mbox{}}  &
${\cal H}^{0}_+(y^\mu)\tau^{2p_2}$
\\
$\frac{H_-}{\Theta^2}$  \parbox[c][8mm]{1mm}{\mbox{}}  &
${\cal H}^{0}_-(y^\mu)\tau^{2p_2}$
\\
$\frac{H_1}{\Theta^2}$  \parbox[c][8mm]{1mm}{\mbox{}}  &
${\cal H}^{0}_1(y^\mu)\tau^{1-p_1}\ln\tau$
\\
$\frac{H_2}{\Theta^2}$  \parbox[c][8mm]{1mm}{\mbox{}}  &
${\cal H}^{0}_2(y^\mu)\tau^{1-p_2}\ln\tau$
\\
$\frac{H_3}{\Theta^2}$ \parbox[c][8mm]{1mm}{\mbox{}}   &
${\cal H}^{0}_3(y^\mu)\tau^{1-p_3}\ln\tau$
\\
\br
\end{tabular}
\end{indented}
\end{table}


\section{Conclusions and Discussion}\label{sec5}
In this work we have considered for the first time the application of
the LWAS in the context of the covariant fluid approach to
cosmology~\cite{COFA,ELVE}.
This produces a system of ODEs for our main variables $\{e_{\alpha}{}^\mu,
\Theta,\sigma_{\alpha\beta}\}$, giving a ``silent'' evolution~\cite{BMP,WAEL}
that, for a collapsing fluid, proceeds towards a Kasner singularity.
However, our analytical results show that at a certain point the
gravito-magnetic field $H_{ab}$ becomes dominant, producing a bounce
to a new Kasner phase and the breakdown of the scheme.  These results
are consistent with the BKL picture for the approach to a generic
spacelike singularity.  On the other hand, for an expanding fluid,
our results show that anisotropies die away and the local
evolution is more and more Robertson-Walker-like.  This confirms
previous analyses~\cite{LADR,BMP,WAEL} and is also in line with recent
investigations on the evolution of non-linearities on superhorizon
scales~\cite{NEW}.

The LWAS is based on the assumption that spatial gradients can be
neglected with respect to time derivatives, whose scale is given by
the Hubble parameter.  Although the approximation scheme is physically
well-motivated it turns out that there is not a well-established
mathematical formulation of the principles on which it is based.
The difficulty in having such a formulation lies in the difficulty
to establish a meaningful way of smoothing a clumpy cosmological
models, which in turn is related to the averaging problem in cosmology,
a problem not yet clarified.

Most of the times the Long-Wavelength Approximation has been directly
applied to the metric components,
and then the spatial gradients of the metric has been neglected,
which of course is a coordinate dependent procedure.  In this work
we have applied it to the quantities of the covariant fluid approach,
which are intrinsically defined once the fluid velocity is prescribed,
but basically the way in which we have implemented it is equivalent
to the previous one.  The reason for this is that we have neglected
completely the spatial curvature tensor ${}^3R_{ab}$, which means
that we have not only neglected spatial derivatives of the connection
components, $\mb{\partial}_\delta\Gamma^{\lambda}_{\alpha\beta}$,
but also the product of them, $\Gamma^\delta_{\alpha\beta}
\Gamma^\epsilon_{\lambda\gamma}$.   The second approximation
basically means that we have neglected the spatial gradients of the
metric.

However, in principle we can generalize this approximation scheme by applying
the idea at the level of the connection instead of applying it at
the level of the metric.  That is, to neglect all the spatial
gradients of the connection, which implies that we will have a
non-vanishing spatial curvature for the first iteration of the scheme.
As one can see, the equations for this case will have essentially
the same form as those of the Bianchi models, which are homogeneous
models with homogeneous spatial curvature.  So we will be still inside
the spirit of the approximation.  The advantage that this generalization
can have is that it would allow us to include the possibility of having
recollapsing regions (regions in which ${}^3R>0$), and in this way one could
be able to explore a more realistic universe model by using the LWAS,
for example studying the evolution, within this approach, of a model
corresponding to a Robertson-Walker universe with initially small
perturbations.
Since the integration functions of the
equations of the LWAS would be space dependent this means that we could have
regions in which the spatial curvature would lead to recollapse, and
other regions that would expand forever.
This would give a picture closer to the idea we have of the universe
from observations, where we have void regions that are in expansion and others
that are accreting matter and collapsing to form cosmic structures.
From a mathematical point of view we expect that the generalized approach
outlined above should be related to the improved LWAS introduced in~\cite{IGIS}
and, for the covariant quantities, should produce a system of ODEs
able to describe the evolution from expansion to recollapse, much like the
equations for ``silent'' models.


\section*{Acknowledgments}
CFS is supported by the EPSRC.


\end{document}